\documentclass[conference]{IEEEtran}
\pdfoutput=1
\IEEEoverridecommandlockouts
\usepackage{cite}
\usepackage{amsmath,amssymb,amsfonts}
\usepackage{algorithmic}
\usepackage{graphicx}
\usepackage{textcomp}
\usepackage{xcolor}
\def\BibTeX{{\rm B\kern-.05em{\sc i\kern-.025em b}\kern-.08em
    T\kern-.1667em\lower.7ex\hbox{E}\kern-.125emX}}

\usepackage[colorinlistoftodos]{todonotes} %
\presetkeys%
{todonotes}%
{inline,backgroundcolor=yellow}{}

\usepackage{csquotes}
\usepackage[bookmarks=false]{hyperref}
\hypersetup{nolinks=false}

\usepackage{tcolorbox}

\usepackage{xspace}
\usepackage{subcaption}

\usepackage{booktabs}

\usepackage{balance}

\usepackage[export]{adjustbox}

\usepackage{soul}

\newcommand{\cc}[1]{\ignorespaces} %
\newcommand{\numRepos}{103\xspace}%
\newcommand{\numFilteredIssues}{10\,459\xspace}%
\newcommand{\numTrainingSetSize}{360\xspace}%

\usepackage[top=0.9in ,foot=-0in, bottom=0.8in, left=0.6in, right=0.6in]{geometry}

\usepackage{fancyhdr}
\usepackage{lastpage}

\begin{document}

\title{A Fault Localization and Debugging Support Framework driven by Bug Tracking Data}

\author{\IEEEauthorblockN{Thomas Hirsch}
\IEEEauthorblockA{\textit{Institute of Software Technology} \\
\textit{Graz University of Technology}\\
Graz, Austria \\
thirsch@ist.tugraz.at}
}

\maketitle

\thispagestyle{fancyplain}
\footnotetext{\textcopyright 2020 IEEE.  Personal use of this material is permitted.  Permission from IEEE must be obtained for all other uses, in any current or future media, including reprinting/republishing this material for advertising or promotional purposes, creating new collective works, for resale or redistribution to servers or lists, or reuse of any copyrighted component of this work in other works.}

\begin{abstract}%
Fault localization has been determined as a major resource factor in the software development life cycle.
Academic fault localization techniques are mostly unknown and unused in professional environments.
Although manual debugging approaches can vary significantly depending on bug type (e.g. memory bugs or semantic bugs), these differences are not reflected in most existing fault localization tools.
Little research has gone into automated identification of bug types to optimize the fault localization process.
Further, existing fault localization techniques leverage on historical data only for augmentation of suspiciousness rankings.
This thesis aims to provide a fault localization framework by combining data from various sources to help developers in the fault localization process.
To achieve this, a bug classification schema is introduced, benchmarks are created, and a novel fault localization method based on historical data is proposed.
\end{abstract}


\section{Introduction}
Debugging is a complex and labour-intensive task, consuming a significant portion of resources during the lifespan of a software product\cite{Ang2017, Tassey2002}.
To tackle this issue a vast number of different approaches from multiple angles have been researched, engineered, and tested.
These approaches range from high level process optimization attempting to lower the number of introduced faults, through the field of automated testing and test creation, to low level tools supporting developers in the debugging process.

Fault localization has been identified as one of the most difficult and time consuming tasks within the debugging process\cite{Vessey1985}.
Over the past decades a plethora of automated Fault Localization (FL) techniques were proposed by researchers (see \cite{WongSurvey2016} for an overview), e.g. Slicing-based, spectrum-based, statistics-based, model-based, data-mining-based, and machine-learning-based approaches.

This thesis started in March 2020, and is rooted in the field of FL techniques.

\emph{Problem statement:}
The effectiveness and performance of these FL techniques has been investigated in detail under laboratory conditions, including comparative evaluation on various benchmarks\cite{Pearson2017, Keller2017}.
However, these advanced debugging techniques are rarely adopted by developers in the field\cite{Parnin2017, Ang2017, Perscheid2017}, with practitioners defaulting to stacktraces, log outputs, symbolic debuggers for FL tasks\cite{Perscheid2017}\cc{Beller2017}. %
The majority of these FL techniques is aiming at the high bar of localizing a fault on statement or file level.
Yet, some academic debugging techniques have been criticized for being based on unrealistic assumptions.
Most notably the \enquote{single-line fault} assumption, the expectation that faults reside in a single incorrect line of code\cite{Lucia2012,Ang2017}, and the \enquote{perfect bug understanding} assumption, forming the premise that the developer can recognize the fault by inspecting the faulty line in isolation. \cite{Ang2017, Parnin2017}\cc{Roszler2012}.
In addition, there is a lack of mature implementations of such FL techniques and the lack of integration into popular IDE's and build systems.

\section{Related work}
\emph{Benchmarks:}
In the past, benchmarking and testing of advanced FL techniques was often performed on artificially created bugs\cite{Pearson2017, Just2014a,Hutchins1994,Do2005}.  %
Researchers voiced concerns that manual or automated seeding of bugs for benchmarking is not adequately reflecting real world scenarios\cite{Boehme2014,Pearson2017,Just2014a}\cc{,Dallmeier2007}.
A growing number of bug benchmarks consisting exclusively of real world bugs is publicly available\cite{Just2014,Lu2005,Bohme2017}\cc{Toth2016,Gyimesi2019,Benton2019,Saha2018,}.
Most of these benchmarks are curated collections for testing specific tool families or specific tools\cite{Amann2016,Dallmeier2007,Boehme2014}\cc{Muvva2020,Soltani2020,Liu2019,Lin2017,Ohira2015,Tomassi2019,,Goues2015,,Rahman2018,Soltani2020,Cifuentes2009,}.
Providing a realistic distribution of different bug types and bug complexities is only in scope of a small number of benchmarks\cite{Just2014,Lu2005,Saha2018}\cc{,Hao2019,,Radu2019,,Riganelli2019}.

\emph{Bug classification:}
There is a plethora of publications regarding bug classification schemas and bug taxonomies\cite{Chillarege92,IEEE2010,Grottke2005a,Tan2014a}\cc{zhou2015,Wan2017Bug,BeizerBoris2001a,Seaman2008,,Ohira2015,}.
The viewpoints under which the bugs are examined, and the intended purpose of the classification can vary widely from schema to schema\cite{Ploski2007}.
Notable examples are Chillarege et al.'s\cite{Chillarege92} Orthogonal Defect Classification (ODC) to aid development process optimization, IEEE's\cite{IEEE2010} Standard Classification for Software Anomalies, Gray's\cite{Gray1985} well-known \enquote{Heisenbugs} and \enquote{Bohrbugs}, and schemas along other dimensions as trigger\cite{Cotroneo2016}, priority\cite{Ortu2016}, severity\cite{Menzies2008}.
A small portion of these schemas are intended to aid investigation and analysis regarding debugging\cite{Ray2014,Tan2014a}\cc{Lu2005,}.

\emph{IR fault localization:}
In contrast to slicing and spectrum based FL methods IR (Information Retrieval) FL methods do not require coverage data. %
Instead they try to leverage on lexical similarities of a bug report with the source code files that are suspected to contain the bug.
Comparison of different IR models, and their suitability for an FL application, has been performed by Lukins et al. and Rao et al.\cite{Lukins2008,Rao2011}.

Further, researchers have investigated augmenting IR based FL with historical project data\cite{Zhou2012,Wang2014a}\cc{,Youm2016}.

\emph{Tool recommendation and combination:}
There is little research into combining multiple FL methods to increase accuracy.
Le et al.\cite{Le2017} presented a method to predict the performance of IR FL methods for a given bug report.
Xuan et al.\cite{Xuan2014} combined suspiciousness rankings of different spectrum based FL methods to increase effectiveness.

\section{Research questions}
The following research questions are carefully crafted to provide answers that are practically relevant and can help to improve FL techniques, and their acceptance in the field.
\begin{tcolorbox}[boxsep=1pt,left=2pt,right=2pt,top=2pt,bottom=2pt]
RQ1 - What types of bugs are occurring in a real world environment, what is their distribution, and which types are the most difficult to fix?
\end{tcolorbox}
Answering RQ1 allows for building realistic benchmarks, while at the same time highlighting the areas where developers would benefit most from additional debugging support.
The bug classification schema created for this task will provide the basis for answering and supporting the following research questions.
\begin{tcolorbox}[boxsep=1pt,left=2pt,right=2pt,top=2pt,bottom=2pt]
RQ2 - Which FL tools and techniques and which manual debugging tools perform best for each type of bug?
\end{tcolorbox}
Results from RQ2 will provide an evaluation of existing FL techniques, and serve as a weighting criteria when combining multiple localization methods.
This information forms the basis for debugging support in the form of automated tool recommendation.
\begin{tcolorbox}[boxsep=1pt,left=2pt,right=2pt,top=2pt,bottom=2pt]
RQ3 - Do certain bug types exhibit specific fault or fix patterns?
\end{tcolorbox}
Determining fault patterns associated with specific bug types enables identification and establishment of relationships to patterns reported by static code checking tools.
The value of these links combined with bug type information for augmenting FL techniques will be investigated.
An example for such a fault pattern associated with resource leaks is a missing \enquote{finally} block where a resource is allocated inside a \enquote{try-catch} block in conjunction with other statements that may raise exceptions.
\begin{tcolorbox}[boxsep=1pt,left=2pt,right=2pt,top=2pt,bottom=2pt]
RQ4 - Can historical data be used as a starting point for FL purposes?
\end{tcolorbox}
To answer RQ4, a novel Machine Learning (ML) based FL approach based on historical data providing variable resolution location information will be implemented.
This approach leverages on existing textual data (e.g. issue tickets, bug reports, commit messages) and corresponding changes in source code.
A model is trained on this data to estimate likelihoods of the fault to reside in a package, class, or method, given a bug report.
Insights gained from RQ2 and RQ3 will be used to augment this localization approach.

\section{Proposed approach}
Following work packages and respective approaches are proposed in order to answer the research questions and reach the targets of this thesis.

\emph{Classification schema (RQ1):}
First step of this thesis is formulating a bug classification schema to categorize bugs based on their root cause.
This schema is then applied as a tool for investigation into the frequencies and resources spent on the various types of bugs.
Furthermore, it will serve as an integral part of the proposed localization techniques and tools.
The classification schema is expected to evolve throughout the duration of this thesis based on preliminary learnings.

\emph{Surveys (RQ1):}
User surveys performed with developers will provide valuable insight into professional environments.
This encompasses the software domain, frequencies of certain bug types, resources spent, applied debugging approaches, and spread of advanced debugging techniques.
Surveying open source projects and manual application of the classification schema can provide further insight.

\emph{Benchmark creation (RQ1):}
A data set is created by mining open source Java projects spanning various software domains.
This benchmark serves as a basis for debugging tool evaluation and data analysis, while providing a realistic distribution of different types and classes of bugs.
The information for each bug consists of a textual bug report, timestamps, all associated commits, commit metadata, Java aware diff statistics, as well as the location of the fix down to method level.
TTF (Time To Fix) and magnitude of code changes are used to estimate difficulty and cost.
The inclusion of bug reports and commit messages supports benchmarking IR based FL techniques.

\emph{Evaluation of existing FL tools and techniques (RQ2):}
Various FL tools and techniques are applied to the benchmark to rank their performance for each bug.
The created data set allows for uncovering connections between bug types, localization tool scores, and effort/difficulty estimates.
This will provide the basis for automated tool recommendation.

\emph{Fault and fix pattern extraction (RQ3):}
Correlation of distinct fault patterns with specific bug types became apparent during manual classification for aforementioned open source study and benchmark creation.
Multiple researchers\cite{Pan2009,Madeiral2018} mined historical bug fix data to identify patterns.
This work will investigate if certain bug types correlate with bug fix patterns and underlying fault patterns.
This mapping is targeted to establishing a link to known anti-patterns and code-smells reported by static code analysis tools as SonarQube\footnote{https://www.sonarqube.org/}. %
The use of identified patterns is manifold, from aiding FL, to extending static code checkers, and to serve as fix templates for automated repair\cite{Lui2019}.

\emph{Novel FL approach (RQ4):}
Some existing IR approaches leverage historical data to augment their suspiciousness rankings.
This thesis investigates the inversion of this hierarchy by identifying a localization starting point based on historical data and then applying other FL methods.
This comprises investigation into the use of historical textual data and corresponding code changes for FL purposes.
The data forming the basis for this approach includes bug tickets, issue tickets, feature requests, pull requests, and commit messages.
ML algorithms are used to create a model reflecting the responsibilities of code components.

Most FL techniques operate on a fixed degree of resolution in the sense of statement, method, or file.
This proposed approach will employ a variable resolution spanning from package to method level.

\emph{Debugging service (RQ4):}
Implementation of an automated debugging service, providing developers with additional information and data, is the final step of this thesis.
This is intended to operate in a fully automated and non intrusive way.
Data from various tools and sources is combined to achieve this.
This includes aforementioned novel debugging approach, automated bug type classification, debugging tool recommendation, IR based localization, and information from static code analysis tools.
Evaluation of this debugging assistant will be performed on the created benchmarks and selected open source projects.

\section{Progress}
A bug classification schema was constructed by adapting and extending upon Tan et al.'s\cite{Tan2014a} bug categorization.
Based on this schema, a user survey was conducted with professional developers to gather information regarding the different bug types\cite{Hirsch2020TODO}.
This included the frequency of those bug types, used debugging tools, times spent on reproducing, locating, and fixing them, as well as the perceived difficulty along those steps.
All bug reports and the corresponding fixes of a medium sized open source application were manually examined.
A similar examination was performed on a subset of the Defects4J\cite{Just2014} benchmark.
The findings reveal that although memory bugs are the rarest, they are the hardest, and most time-consuming to reproduce and locate.
Concurrency bugs falling in second in terms of time consumption and difficulty.
Semantic bugs are the most common, their mean and median time to fix is much smaller, most notably the mean and median time to reproduce and locate is significantly smaller than for memory and concurrency bugs.

We created a bug benchmark by mining open source projects hosted on Github.
As of writing this, the collection contains \numFilteredIssues bugs from \numRepos different open source Java projects.
Researchers have already demonstrated the feasibility of automated classification using ML schemas\cite{Thung2012,Ray2014,Li2006a} on various bug categorization schemas\cc{Hernandez-Gonzalez2018,Lopes2020}.
To automate the classification process for our bug schema, we applied Natural Language Processing (NLP) in conjunction with ML algorithms\cite{Hirsch2020WorkshopTODO}.
Training set creation is currently ongoing.
Preliminary testing with a training set size of \numTrainingSetSize shows promising results, reaching mean F1 scores of 74\,\%.

\section{Threats to validity}
As of now, data mined from open source Java projects form the basis for the benchmark and performed experiments.
The programming language and software domain have an influence on distribution of bug types, fault patterns, and debugging strategies, approaches, and tools.
It is planned to extend the scope of this research onto at least two more object-oriented languages to illustrate the validity of the approach on a wider scope.
Further, including a wide variety of different software domains enables investigation into biases and cross validation.

One of the biggest threats is introduced by mining data exclusively from open source projects.
An adequately sized data corpus based on proprietary software is not available at this time.
Applying a list of project selection criteria is intended to lessen possible biases.
Examples of the selection criteria are: the project is driven by a well-known organization, the size of the project, spread, and popularity of the software.

The ML localization approach depends strongly on the amount of historical data that is available in a software project.
This creates a bootstrapping problem and inhibits application on projects with small bodies of historical data.
This thesis will include investigation into these effects and limiting factors, and try to identify thresholds.
Furthermore, the proposed localization approach assumes that the code base is stable in the sense of responsibility of components and their location.
However, in an actively developed and maintained project responsibilities will be in flow due to feature introduction, refactorings and re-implementations.
The influence of the project's stability on the localization approach will be investigated.

\section{Milestones}
\emph{2020:}
Development of the bug classification schema and evaluation of said schema by performing user studies and surveying open source projects.
Creating a bug benchmark from Java projects.

\emph{2021:}
Evaluating existing FL tools on the benchmark and investigating the relation of bug types and tool performance.
Mining fault and fix patterns from the benchmark, and organizing a list of patterns known to checking tools with the goal of establishing links between them.
Implementing a novel ML FL approach based on historical data.

\emph{2022:}
Extending the benchmarks to other programming languages including fault pattern mining and alignment of static checkers for this language.
Implementing an automated FL service combining the novel approach, fault pattern information, and existing fault localization tools.

\emph{2023:}
Defending the thesis.

\section{Expected contribution}
The expected contributions of this thesis include a bug classification schema and taxonomy to benefit debugging efforts, as well as bug benchmarks based on realistic distributions of bug types.
These benchmarks are usable for a wide variety of FL tools including IR based fault localization.
Further, this research will yield a novel ML based FL method leveraging on historical, and implementation data, providing location information in a variable resolution.
This will include an implementation of a debugging assistant combining the ML based localization approach with existing FL techniques and information from static checkers to provide location information and tool recommendation.

\section{Acknowledgments}
This thesis is supervised by professor Franz Wotawa at the Technical University of Graz at the department for software technologies.
I want to thank professor Franz Wotawa and Birgit Hofer for the exceptional support.
This thesis is part of a project funded by the Austrian Science Fund (FWF) under contract number P 32653.

\bibliographystyle{IEEEtran}

\bibliography{library}

\end{document}